\documentclass[%
aip,
 amsmath,amssymb,
preprint,%
]{revtex4-1}

\usepackage{graphicx}
\usepackage{dcolumn}
\usepackage{bm}

\usepackage[utf8]{inputenc}
\usepackage[T1]{fontenc}
\usepackage{mathptmx}
\usepackage{etoolbox}

\usepackage{color}

\makeatletter
\def\@email#1#2{%
 \endgroup
 \patchcmd{\titleblock@produce}
  {\frontmatter@RRAPformat}
  {\frontmatter@RRAPformat{\produce@RRAP{*#1\href{mailto:#2}{#2}}}\frontmatter@RRAPformat}
  {}{}
}%
\makeatother
\begin{document}


\title[Reversible vertical positioning of acoustically levitated particle using a spiral reflector]{Reversible vertical positioning of acoustically levitated particle using a spiral reflector}

\author{Yusuke Koroyasu}
 \email{koroyu@digitalnature.slis.tsukuba.ac.jp}
 \affiliation{Graduate School of Comprehensive Human Science, University of Tsukuba, Tsukuba, Ibaraki 305-8550, Japan}

\author{Takayuki Hoshi}
 \affiliation{Pixie Dust Technologies, Inc., Chuo-ku, Tokyo 104-0028, Japan}

\author{Yoshiki Nagatani}
 \affiliation{Pixie Dust Technologies, Inc., Chuo-ku, Tokyo 104-0028, Japan}

\author{Daichi Tagami}
 \affiliation{Graduate School of Comprehensive Human Science, University of Tsukuba, Tsukuba, Ibaraki 305-8550, Japan}

\author{Yoichi Ochiai}
 \affiliation{Pixie Dust Technologies, Inc., Chuo-ku, Tokyo 104-0028, Japan}
 \affiliation{R\&D Center for Digital Nature, University of Tsukuba, Tsukuba, Ibaraki 305-8550, Japan}
 \affiliation{Institute of Library, Information and Media Science, University of Tsukuba, Tsukuba, Ibaraki 305-9550, Japan}
 \affiliation{Tsukuba Institute for Advanced Research (TIAR), University of Tsukuba, Tsukuba, Ibaraki 305-8577, Japan}

 \author{Tatsuki Fushimi}
 \affiliation{R\&D Center for Digital Nature, University of Tsukuba, Tsukuba, Ibaraki 305-8550, Japan}
 \affiliation{Institute of Library, Information and Media Science, University of Tsukuba, Tsukuba, Ibaraki 305-9550, Japan}
 \affiliation{Tsukuba Institute for Advanced Research (TIAR), University of Tsukuba, Tsukuba, Ibaraki 305-8577, Japan}

\date{\today}

\begin{abstract}
Dynamic positioning in acoustic levitation typically depends on active control of the transducers phases, which necessitates complex driving electronics. While mechanically actuated reflectors offer a simpler alternative, achieving reversible transport along the vertical axis solely through mechanical actuation remains challenging. Here, we demonstrate vertical particle translation using a rotating spiral reflector with a half-wavelength pitch. With the rotation axis laterally offset relative to the acoustic focus, the spiral surface functions as a series of translating slopes. Experimental and numerical results confirm stable, bidirectional transport, yielding a vertical displacement of approximately $0.58\lambda$ per revolution and a maximum height of $3.18\lambda$, with radial confinement maintained within $0.24\lambda$. This approach provides a cost-effective solution for non-contact sample handling without active phase control.
\end{abstract}

\maketitle

Mid-air acoustic levitation is a technique that utilizes acoustic radiation forces to suspend and manipulate objects in the air without physical contact. This capability enables a wide range of applications, including analytical chemistry~\cite{santesson2004airborne,westphall2008mass}, materials science~\cite{trinh1985compact,weber2012acoustic,drewitt2024mightylev}, and biomedical research~\cite{vasileiou2016toward,puskar2007raman,sundvik2015effects}, and laboratory automation~\cite{Foresti2013a,andrade2018automatic,watanabe2018contactless}, as well as additive manufacturing~\cite{ezcurdia2022leviprint,chen2025acoustics,chen2025omnidirectional}. A standard single-axis configuration establishes a standing wave field between ultrasonic emitters and an opposing reflector~\cite{xie2001parametric,andrade2015particle,qin2022acoustic}, or alternatively between two opposed emitters~\cite{marzo2017tinylev,drewitt2024mightylev,ochiai2014three}. Within the standing wave field, particles are trapped at pressure nodes located at half-wavelength intervals along the vertical axis.

Dynamic positioning of acoustically levitated objects is generally achieved through two distinct methodologies: modulation of the sound source or mechanical actuation of a reflector. The former typically utilizes phased array transducers to synthesize diverse acoustic fields by dynamically regulating the phase delay and amplitude of individual emitting transducers~\cite{hoshi2010noncontact,marzo2017ultraino,morales2021generating,montano2023openmpd,kondo2021mid,fushimi2019acoustophoretic,hirayama2019volumetric,ochiai2014three}. Although phased array transducers enable complex field synthesis, they require costly multichannel driving circuits, which increases hardware complexity~\cite{marzo2017tinylev}. The latter method reconfigures the standing wave field through mechanical actuation of the reflector~\cite{andrade2015particle,qin2022acoustic}. This mechanical approach eliminates the need for complex electronic control systems, thereby offering a cost-effective solution suitable for practical laboratory and industrial applications.

For example, Qin \textit{et al.} demonstrated particle transport using a screw-shaped reflector with a quarter-wavelength pitch, placed coaxially with a flat emitter~\cite{qin2022acoustic}. In this configuration, vertically translating the reflector creates an azimuthally dependent nodal pattern, guiding particles along helical trajectories on a conical surface. While this demonstrates the feasibility of mechanical reflector actuation, the transport mechanism is inherently irreversible and couples radial and vertical displacements. Consequently, a method for reversible positioning strictly along the vertical axis, driven by simple mechanical rotation, has not yet been realized.

\begin{figure}[t]
    \centering
    \includegraphics[width=0.50\textwidth]{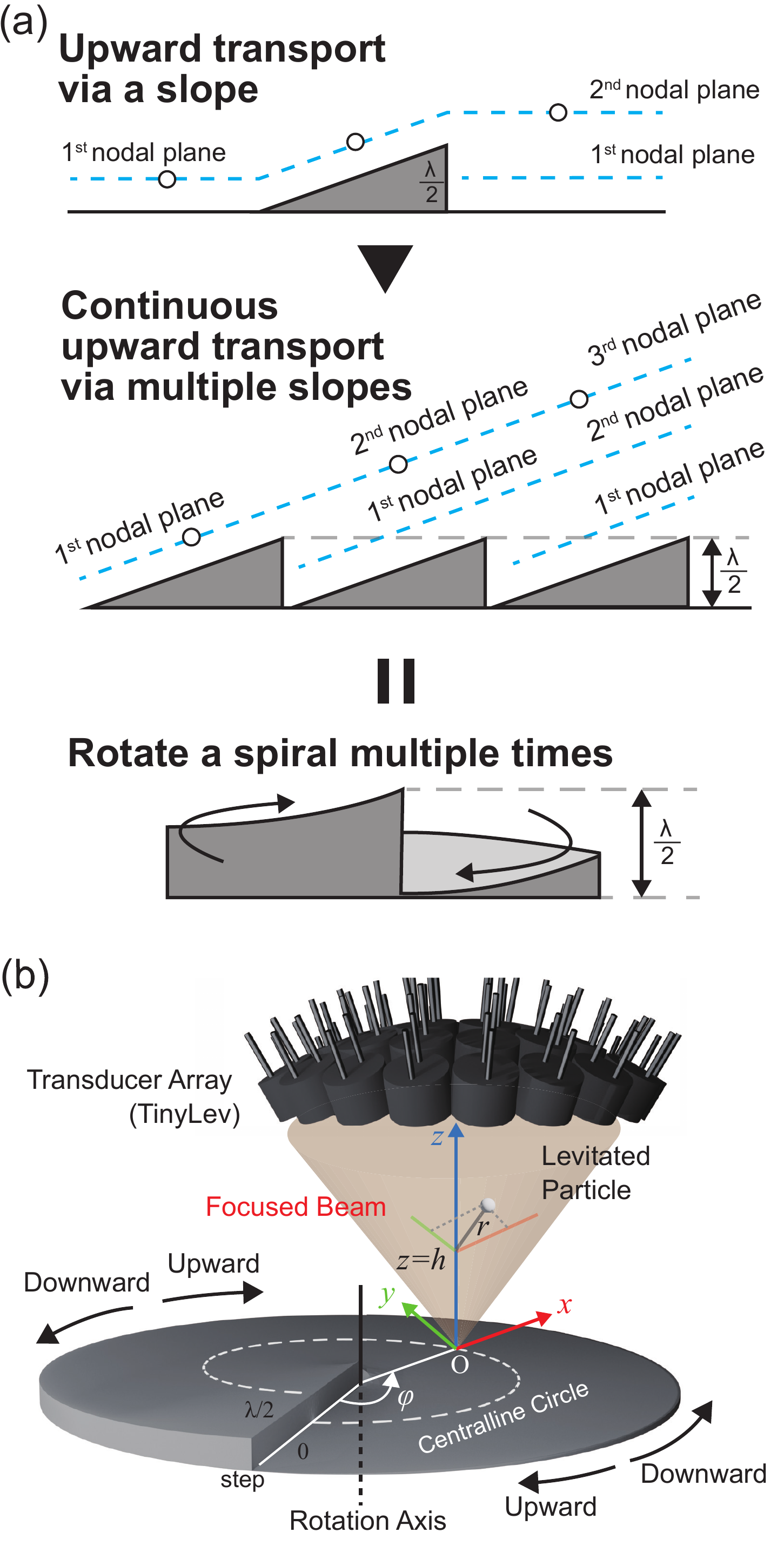}
\caption{Proposed vertical transport mechanism and experimental setup. (a) Conceptual schematic: A slope translating beneath the particle induces a continuous nodal ascent (top). This principle is extended to a periodic sequence (middle) and mapped onto a rotational coordinate system (bottom) to realize continuous transport within a compact footprint. (b) Experimental geometry: The origin $O$ corresponds to the acoustic focus of the transducer array. The $x$- and $y$-axes are defined as the normal and tangential directions to the reflector's centerline circle (white dashed circle), respectively, defining the radial distance from the vertical axis as $r = \sqrt{x^2+y^2}$. The rotation axis of the spiral reflector is laterally offset by $x = -25$ mm to align the incident beam with the centerline circle. The reflector surface height $H(\varphi)$ increases linearly with the rotation angle $\varphi$, with a helical pitch of $\frac{\lambda}{2}$.}
    \label{fig:schematic}
\end{figure}

\begin{figure*}[t]
    \centering
    \includegraphics[width=0.95\textwidth]{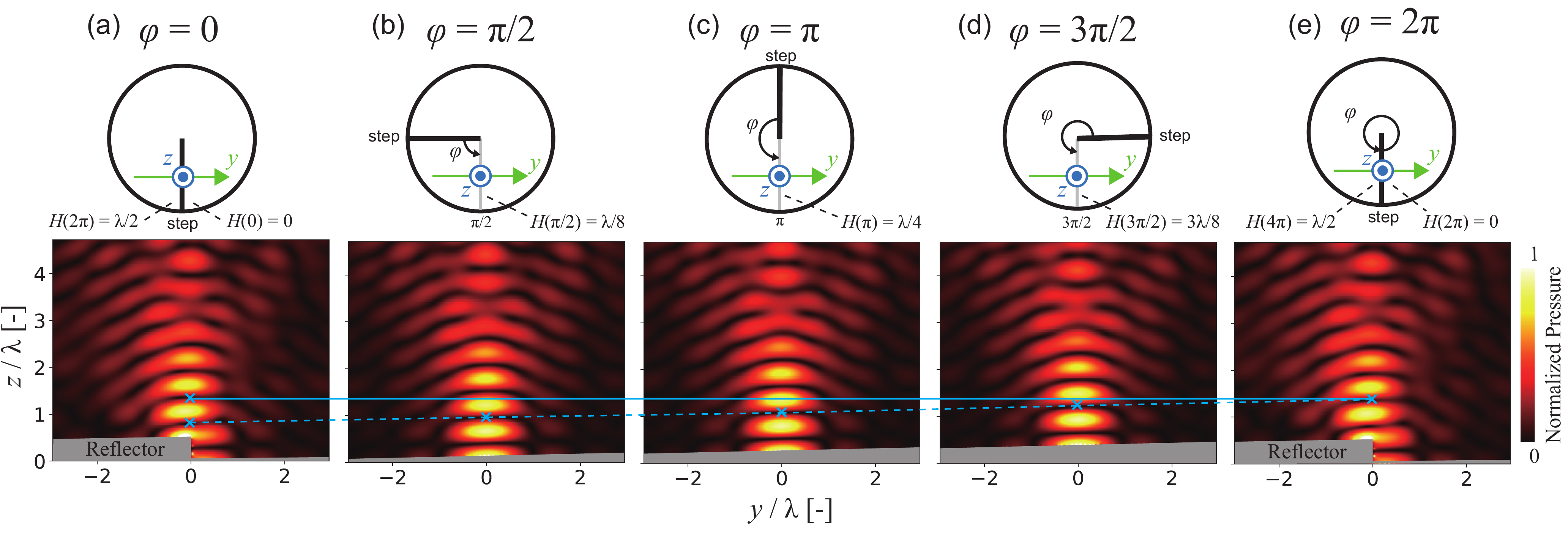}
    \caption{Numerical demonstration of continuous vertical transport over one full revolution. (a)--(e) Normalized acoustic pressure distributions in the $y$--$z$ plane calculated using the boundary element method (BEM) at rotation angles $\varphi = 0, \frac{\pi}{2}, \pi, \frac{3\pi}{2},$ and $2\pi$, respectively. Blue crosses mark pressure nodes along the vertical axis. The dashed cyan line traces the upward trajectory of a transitioning node. The solid cyan line connects the node at $\varphi=2\pi$ to the node at $\varphi=0$ located at the same vertical height, demonstrating that the particle is seamlessly handed over to the subsequent nodal cycle.}
    \label{fig:numerical}
\end{figure*}

\begin{figure*}[t]
    \centering
    \includegraphics[width=0.85\textwidth]{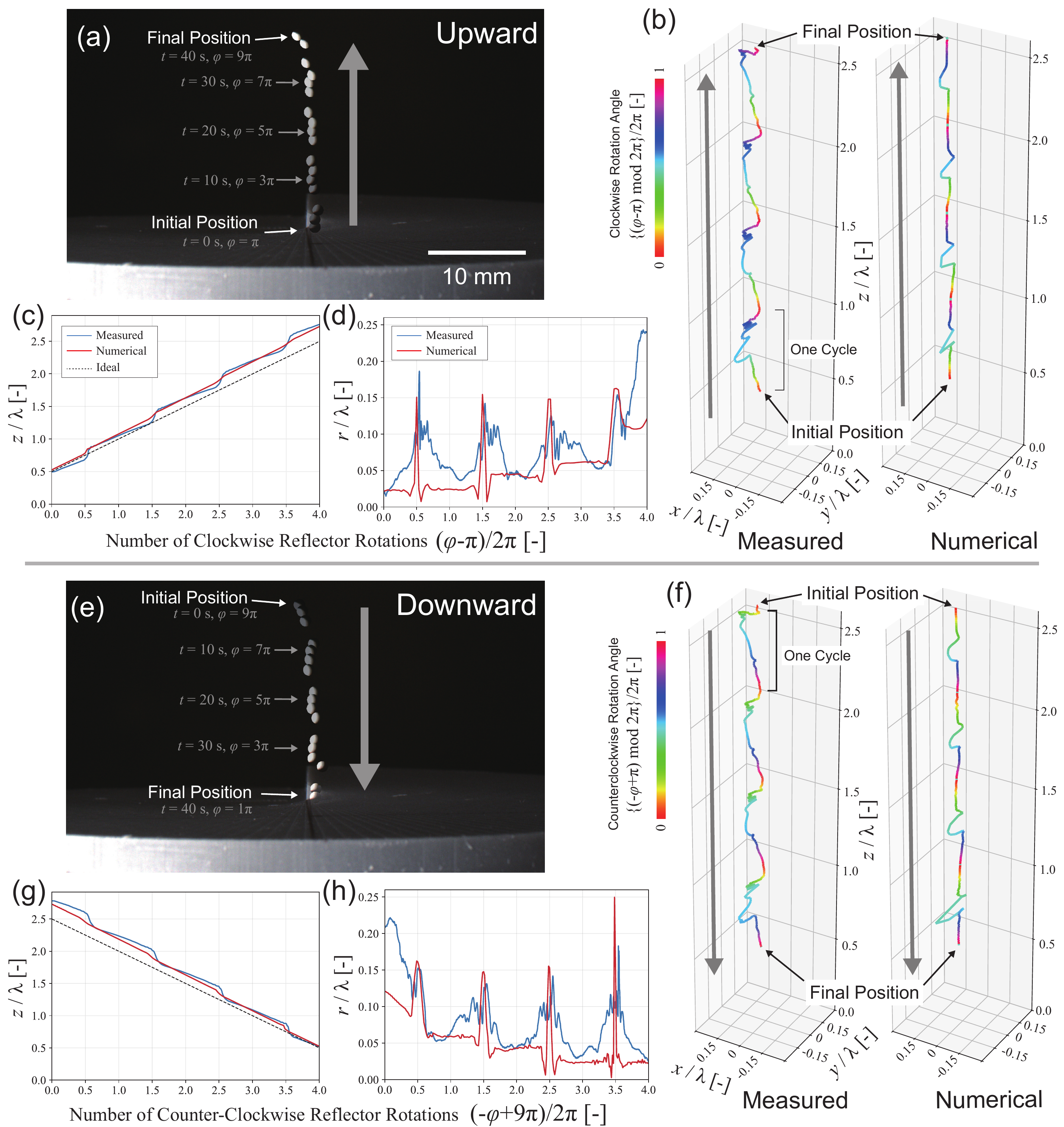}
    \caption{Experimental validation and numerical analysis of reversible vertical transport. (a)--(d) Upward transport induced by clockwise rotation of the spiral reflector (from $\varphi=\pi$ to $9\pi$) and (e)--(h) downward transport induced by counter-clockwise rotation (from $\varphi=9\pi$ to $\pi$). (a), (e) Composite images visualizing the vertical motion. The particle images were extracted from individual frames and superimposed onto the background, with brightness linearly increased from 30\% at the initial position to 100\% at the final position to indicate the time evolution. (b), (f) Comparison of three-dimensional trajectories between experimental measurements and BEM simulations. The color gradient represents the rotation angle within one cycle. (c), (g) Normalized vertical position $\frac{z}{\lambda}$ as a function of the number of reflector rotations. The black dashed lines indicate the ideal vertical displacement ($\frac{\lambda}{2}$) per revolution. (d), (h) Normalized radial displacement $\frac{r}{\lambda}$ from the vertical axis. The multimedia view is a demonstration of upward and downward transport videos (Multimedia view).}
    \label{fig:ex_vs_numerical}
\end{figure*}

In this Letter, we propose vertical particle translation using a rotating spiral reflector. As shown in Fig.~\ref{fig:schematic}, the reflector is defined by a helical surface profile $H(\varphi)$ with a half-wavelength pitch:
\begin{equation}
H(\varphi) = \frac{\lambda}{4\pi}\,(\varphi \bmod 2\pi).
\end{equation}
Unlike the coaxial arrangement of Qin \textit{et al.}~\cite{qin2022acoustic} which induces helical trajectories, our setup laterally offsets the rotation axis from the acoustic focus (Fig.~\ref{fig:schematic}(b)). This offset enables the rotating spiral reflector to function as a series of translating slopes, converting mechanical rotation directly into bidirectional vertical transport.

The fundamental design concept is illustrated in Fig.~\ref{fig:schematic}(a). Since pressure nodes are spatially locked to the reflector, horizontally translating a slope induces vertical trap displacement. Consequently, a particle levitated at the $n$-th node is transported to the $(n+1)$-th node as a slope of height $\frac{\lambda}{2}$ passes beneath it~\cite{hirayama2022high}. While arranging such slopes in series enables continuous transport, the linear configuration necessitates an excessive lateral distance. To achieve a space-efficient design, we mapped the geometry onto a rotating spiral. Here, the step height is set to $\frac{\lambda}{2}$, ensuring that the pressure node connects seamlessly across the geometric discontinuity.

Figure~\ref{fig:schematic}(b) shows the experimental apparatus. The acoustic field was generated by a single-sided array of 36 ultrasonic transducers (UT1007-Z325R, SPL) based on the TinyLev design~\cite{marzo2017tinylev}. The transducers were driven by 40~kHz square-wave signals generated by an Arduino Nano and amplified to $24~\mathrm{V_{\mathrm{pp}}}$ via dual full-bridge drivers (L298N). The array was positioned 60~mm above the reflector, aligning its geometric focus with the base height. The rotation axis was laterally offset by 25~mm from the acoustic focus. The reflector, 100~mm in diameter, was fabricated using stereolithography (Form 4, Formlabs Inc.) and mounted on a motorized rotation stage (PRMTZ8/M, Thorlabs Inc.). The levitated particle was an expanded polystyrene (EPS) sphere ($a=0.78$~mm, $\rho_p=32.4~\mathrm{kg\,m^{-3}}$), characterized with a digital microscope and an ultramicrobalance (MSA2.7S-000-DM, Sartorius). Two orthogonal industrial cameras (BFS-U3-16S2C-CS, Teledyne FLIR), equipped with lenses with focal lengths of 12~mm (FL-BC1220-9M, Optowl) and 12.5~mm (LM12HC, Kowa), recorded particle motion in the $xz$ and $yz$ planes at 30~fps for trajectory reconstruction. In each frame, the particle centroid was extracted via background subtraction and the OpenCV function \texttt{minEnclosingCircle}.

Pressure fields were calculated using the boundary element method (BEM)~\cite{hirayama2022high} via the open-source library AcousTools~\cite{mukherjee2025acoustools}. Transducers were modeled as circular pistons, and the reflector as a sound-hard boundary discretized into a triangular mesh with an element size of $\approx \frac{\lambda}{10}$. The acoustic medium was characterized by the standard properties of air at room temperature: sound speed $c_0 = 341~\mathrm{m\,s^{-1}}$ and density $\rho_0 = 1.21~\mathrm{kg\,m^{-3}}$.

Figure~\ref{fig:numerical} shows the variation of the numerically calculated pressure field with rotation angle $\varphi$. As the reflector rotates from $\varphi=0$ to $2\pi$ (Figs.~\ref{fig:numerical}(a)--(e)), the rising spiral surface at the vertical axis drives the pressure nodes upward along the trajectory (dashed cyan line). At $\varphi=2\pi$, the nodes are shifted vertically by $\frac{\lambda}{2}$, coinciding with the adjacent node at $\varphi=0$ (solid cyan line). This periodicity ensures seamless particle transport across the geometric step.

To numerically simulate the particle trajectory, we employed the Gor'kov potential $U$ within the Rayleigh regime ($a \ll \lambda$)~\cite{gorkov1961dokl, bruus2012acoustofluidics}. The time-averaged potential is expressed as:
\begin{equation}
U = V_p \left( \frac{1}{2} \kappa_0 f_1 \langle p^2 \rangle - \frac{3}{4} \rho_0 f_2 \langle \boldsymbol{v}^2 \rangle \right),
\label{eq:gorkov_potential}
\end{equation}
where $V_p=\frac{4}{3}\pi a^3$ is the particle volume, $\boldsymbol{v} = \frac{\nabla p}{j\omega\rho_0}$ is the acoustic particle velocity, and $\langle \cdot \rangle$ denotes time averaging. The monopole and dipole scattering coefficients are defined as $f_1 = 1 - \frac{\kappa_p}{\kappa_0}$ and $f_2 = \frac{2(\rho_p - \rho_0)}{2\rho_p + \rho_0}$, respectively, where $\rho$ and $\kappa$ represent density and compressibility, with subscripts $0$ and $p$ denoting the medium and the particle. The particle trajectory induced by the reflector rotation was reconstructed by tracking the sequence of equilibrium positions. Assuming the particle remains trapped at the local potential minimum, the position at each rotation angle $\varphi$ was determined by numerically minimizing the three-dimensional potential field $U$ using the L-BFGS-B algorithm (SciPy library). To simulate continuous motion, $\varphi$ was incremented in steps of $12^{\circ}$, where the equilibrium position from the previous step served as the initial guess for the subsequent minimization.

To experimentally demonstrate upward transport, a particle was initially levitated at the pressure node closest to the reflector surface ($z \approx \frac{\lambda}{2}$). The reflector was rotated clockwise at a low speed of $6~\mathrm{rpm}$ for four complete revolutions ($\varphi=\pi$ to $9\pi$). Figure~\ref{fig:ex_vs_numerical}(a) presents a composite image of the levitated particle, visualizing the continuous upward motion (Multimedia view). The reconstructed three-dimensional trajectory is compared with the numerical prediction in Fig.~\ref{fig:ex_vs_numerical}(b). The experimental data show good agreement with the numerical results, confirming that the particle is stably transported vertically while remaining confined near the central axis.

Figures~\ref{fig:ex_vs_numerical}(c) and (d) present the vertical ($z$) and radial ($r=\sqrt{x^2+y^2}$) displacements as a function of the number of reflector rotations. As shown in Fig.~\ref{fig:ex_vs_numerical}(c), the particle moves upward monotonically from the initial position. The experimental vertical transport rate of $0.58\lambda$ per revolution agrees with the numerical result of $0.55\lambda$; both values exceed the ideal geometric pitch of $0.50\lambda$, indicating a deviation from the geometric prediction. In contrast to the nearly linear numerical prediction, the experimental vertical trajectory exhibits periodic fluctuations at half-integer intervals (i.e., at 0.5, 1.5, 2.5, and 3.5 cycles). These fluctuations coincide with the timing of the particle passing over the reflector steps, where transient peaks also appear in the radial displacement (Fig.~\ref{fig:ex_vs_numerical}(d)). Although the numerical model predicts sharp peaks for these perturbations, the experimental radial response is notably smoother. At higher heights, a gradual radial drift is observed; this behavior is consistent with the numerical results and is attributed to the inherent off-axis deviation of the pressure nodes (Fig.~\ref{fig:numerical}). Crucially, despite these factors, the maximum radial displacement remains within $0.24\lambda$, confirming effective confinement near the vertical axis.

To investigate reversibility, the reflector was rotated counter-clockwise ($6~\mathrm{rpm}$) from $\varphi=9\pi$ back to $\varphi=\pi$. Figures~\ref{fig:ex_vs_numerical}(e)--(h) present the downward transport results (Multimedia view). The composite image (Fig.~\ref{fig:ex_vs_numerical}(e)) and the three-dimensional trajectory (Fig.~\ref{fig:ex_vs_numerical}(f)) reveal that the particle follows a downward path mirroring the upward trajectory. The vertical displacement (Fig.~\ref{fig:ex_vs_numerical}(g)) confirms that the downward motion is monotonic and symmetric to the upward motion. The experimental downward rate of $-0.58\lambda$ per revolution agrees with the numerical value of $-0.55\lambda$; these magnitudes match those observed during the upward transport. The radial displacement (Fig.~\ref{fig:ex_vs_numerical}(h)) shows that, despite transient deviations at half-integer intervals, the particle remains confined near the vertical axis with a maximum radial displacement of $0.22\lambda$. These results demonstrate that the spiral reflector enables stable, bidirectional vertical positioning driven solely by mechanical rotation.

The vertical transport range is primarily limited by the degradation of trap quality at higher vertical positions. In contrast to the demonstration in Fig.~\ref{fig:ex_vs_numerical}, we quantified the operational limit by performing five upward transport trials driven by clockwise rotation of the reflector until the particle was ejected. In these trials, the particle consistently reached a maximum height of $3.18\lambda$ (standard deviation $< 0.01\lambda$). To investigate trap stability in the absence of rotation, we conducted stationary levitation tests with the reflector fixed at $\varphi=\pi$. These tests revealed that stable trapping was unattainable beyond the fifth pressure node ($z \approx 2.50\lambda$), even with manual placement. These results confirm that the transport upper bound is dictated by the distortion of the standing wave field (Fig.~\ref{fig:numerical}(c)), which shifts pressure nodes off-axis, resulting in insufficient trapping force to counteract gravity or withstand disturbances. This distortion is attributed to the short focal length (60~mm) of the emitter array used in this study (TinyLev~\cite{marzo2017tinylev}). Extending the vertical transport range thus requires an emitter array with a longer focal length placed at a greater distance to maintain pressure nodes on the vertical axis at higher positions.

The geometric design of the reflector presents a trade-off between device compactness and transport stability. In this study, a reflector radius of $50~\mathrm{mm}$ ($5.87\lambda$) was adopted. However, reducing this radius introduces two destabilizing mechanisms. First, as the reflector size approaches the incident beam width, it induces unwanted reflections that disrupt the acoustic trap. Second, decreasing the radius necessitates a steeper spiral slope to maintain the fixed vertical pitch of $\frac{\lambda}{2}$. This steeper inclination increases the angular deviation of the reflected wave relative to the vertical axis, resulting in a radial shift of the pressure nodes. Maintaining a sufficient reflector radius is thus crucial to prevent such off-axis deviations and ensure stable confinement.

The transport dynamics depend on the reflector pitch relative to the standing wave periodicity. In this study, a pitch of $\frac{\lambda}{2}$ was selected to match the inter-nodal spacing, yielding a vertical displacement of approximately $0.58\lambda$ per revolution. Although increasing the pitch to integer multiples of the half-wavelength (e.g., $1\lambda$) would theoretically increase transport speed, the increased step height induces scattering that disrupts nodal continuity, thereby compromising particle stability. Moreover, larger steps degrade vertical positioning resolution. Therefore, the $\frac{\lambda}{2}$ pitch is a conservative design choice that prioritizes stable transport and precise control.

In conclusion, we have demonstrated a mechanical method for continuous and reversible vertical particle positioning using a rotating spiral reflector with a helical pitch of $\frac{\lambda}{2}$. By laterally offsetting the rotation axis, this configuration enables linear transport along the vertical axis through mechanical rotation of the reflector. Experimental validation, supported by numerical simulations, confirmed stable bidirectional transport with radial confinement. These findings present a cost-effective alternative to complex electronic phase control, providing a simplified solution for single-axis acoustic manipulation.

\begin{acknowledgments}
This project was funded by JSPS KAKENHI (Grant Number 17H04688). The authors declare no conflicts of interest associated with this manuscript. We used OpenAI’s GPT-5.1/5.2, and Google’s Gemini 3 Pro for light language editing.
\end{acknowledgments}

\section*{Data Availability Statement}
The data that support the findings of this study are openly available in Zenodo at XXXXXX.

\nocite{*}
\bibliography{mybib}

\end{document}